# MFU-type metal-organic frameworks as host materials of confined supercooled liquids

J. K. H. Fischer,[1] P. Sippel,[1] D. Denysenko,[2] P. Lunkenheimer,[1,a)] D. Volkmer,[2] and A. Loidl[1]

[1]*Experimental Physics V, Center for Electronic Correlations and Magnetism, University of Augsburg, 86135 Augsburg, Germany*
[2]*Chair of Solid State and Material Science, Institute of Physics, University of Augsburg, 86135 Augsburg, Germany*

In this work we examine the use of metal-organic framework (MOF) systems as host materials for the investigation of glassy dynamics in confined geometry. We investigate the confinement of the molecular glass former glycerol in three MFU-type MOFs with different pore sizes and study the dynamics of the confined liquid via dielectric spectroscopy. In accord with previous reports on confined glass formers, we find different degrees of deviations from bulk behavior depending on pore size, demonstrating that MOFs are well-suited host systems for confinement investigations.

a) Electronic mail: peter.lunkenheimer@physik.uni-augsburg.de

## I. INTRODUCTION

The continuous slowing down of molecular motions, when a low-viscosity liquid is supercooled and finally transforms into a glass, is still only fairly understood from a microscopic point of view.[1,2,3,4] To explain the super-Arrhenius temperature dependence of the relaxational dynamics at the glass transition, an increasingly cooperative nature of molecular motions at low temperatures is often invoked.[2,5,6,7,8,9,10] To learn more about cooperativity and the glass transition in general, the investigation of supercooled liquids that are confined in spaces of nanometer size has proven a very useful tool.[11,12,13,14,15,16,17,18,19,20,21] For example, as soon as the cooperativity length scale exceeds that of the confining geometry, clear deviations from bulk behavior are expected. Confinement measurements are also of interest as they often allow for an effective suppression of crystallization, thus enabling the investigation of materials that are difficult or impossible to supercool in bulk form. The most prominent example is water, whose glass temperature and suspected fragile-to-strong transition lie in the so-called "no-man's land" between about 150 and 235 K, where crystallization precludes its investigation in bulk form.[22,23,24,25]

Various host materials have been used to provide an environment with confined geometry for glass-forming liquids.[16,17,19,20] This includes amorphous materials such as the so-called "controlled pore glasses", with relatively well-defined pore sizes larger than 2.5 nm, and materials prepared by sol-gel techniques like silica xerogel or aerogel, which, however, have a broad distribution of pore sizes.[20] Zeolites, silicates having rather small pore sizes up to about 1.3 nm (Ref. 20), are examples for crystalline porous materials that are often employed for confinement investigations. Another commonly used silica-based material is MCM-41, which can be prepared with pore sizes varying between about 1.6 and 10 nm.[26] Several other materials also are available (for an overview, see Ref. 20). It should be noted that in many of the above-mentioned cases, the aspect ratios of the pores are far from one. For example, in zeolites ratios of $10^4$ and in MCM-41 ratios of $10^2$ - $10^3$ are found.[20] Thus, liquids confined in these materials can be considered as essentially one-dimensional systems. In Ref. 16 it was shown that the properties of confined liquids can critically depend on dimensionality.

In the present work, we examine the possible use of so-called metal-organic frameworks (MOFs) as host materials for confined supercooled liquids. These materials comprise metal ions or clusters (so-called secondary building units) that are joined together by bridging organic ligands (linkers), thereby forming extended three-dimensional crystalline frameworks with significant porosity.[27,28,29,30,31,32] Until now, the tremendous interest in this material class is mainly triggered by the many promising functionalities of MOFs such as catalysis or the storage of gaseous fuels.[29,30,31,32,33,34,35,36,37] However, even though more than 20000 MOFs are known,[32] providing a larger variety than any other class of porous materials, to our knowledge until now there are no studies of supercooled liquids confined in MOFs.

The pore sizes of MOFs range between about 0.4 and 10 nm.[30,32,33] Of special interest is the region 1 - 4 nm, where various MOFs are available and which is believed to be the characteristic cooperativity length scale of supercooled liquids.[8,38,39] In particular, the region of 1 - 2 nm is relatively difficult to access with other materials,[33] especially if requiring 3D confinement. While in many other host systems the pore dimensions are considerably distributed,[20,40] various MOFs are available where the pore sizes are well-defined. Moreover, the pore sizes of MOFs and the apertures between the pores can be varied, e.g., by using different organic linkers. Exchanging the linkers and/or secondary building units also enables the tuning of the interactions between guest molecules and pore walls, which play an important role in the interpretation of confinement measurements.

In the present work, we demonstrate the feasibility of confinement investigations of glassy dynamics and the glass transition using MOFs as host material. For this purpose, we provide broadband dielectric measurements of glycerol, one of the most investigated glass formers, confined in three MOFs, MFU-1, MFU-4, and MFU-4*l*, where MFU stands for "Metal-Organic Framework Ulm-University".[36,41,42] Their relevant pore sizes range between 1.2 and 1.9 nm. The structural $\alpha$-relaxation process of glycerol is clearly identified and its variation upon confinement is investigated in detail.

## II. HOST MATERIALS

MFU-1 is a cobalt-containing MOF which crystallizes in the cubic crystal system, in the space group $P\bar{4}3m$, and contains pores with a maximum diameter of 1.81 nm, which are connected by 0.9 nm apertures to each other (cf. Fig. 1).[36,37] Due to its large pores, MFU-1 can be readily saturated with glycerol via vapor diffusion from the gas phase (as described in section III). Estimation with the PLATON/SQUEESE program[43] reveals a micropore volume of 1.49 cm$^3$g$^{-1}$ (65.7 % of the unit cell volume). However, the pore volume determined experimentally from argon sorption isotherm is considerably lower (0.57 cm$^3$g$^{-1}$), which is due to partial interdigitation of the framework, leading to the reduced solvent uptake capacity.[37] Thus, the experimentally determined average number of glycerol molecules per unit cell $N_{\text{exp}}$ is about 4.1, which has to be compared to a calculated value $N_{\text{sim}}$ of 17 molecules per unit cell for the non-interdigitated structural model of MFU-1 (Table 1). The actual number of glycerol molecules per pore in our samples certainly is larger than 4.1 because only 38 % of the pore volume is freely accessible as revealed by the mentioned argon-sorption experiments. However, available experimental data do not allow determining the distribution of interdigitating fragments in the framework and thus no exact number of glycerol molecules per pore can be given.

In contrast, interdigitation cannot occur in samples of MFU-4. MFU-4 is a zinc-containing metal-organic framework which crystallizes in the cubic crystal system, in the space group $Fm\bar{3}m$, and contains pores with a maximum diameter of 1.19 nm, which are connected by 0.25 nm apertures to each other (cf. Fig. 1).[41] Owing to the very small limiting diameter of the pore apertures in this framework, an alternative procedure for the saturation of MFU-4 crystal specimen with glycerol (heating with liquid glycerol at 140 °C) was required (see section III).

Table I. Pore sizes, simulated and experimentally found numbers of glycerol molecules per formula unit of MFU-1, MFU-4, and MFU-4*l*. (It should be noted that $N_{\text{sim}}$ or $N_{\text{exp}}$ do not correspond to the number of molecules per pore.)

| MOF | composition | pore size (nm) | $N_{\text{exp}}$ | $N_{\text{sim}}$ |
|---|---|---|---|---|
| MFU-1 | C$_{48}$H$_{48}$N$_{12}$OCo$_4$ | 1.81 | 4.1 | 17 |
| MFU-4 | C$_{18}$H$_6$Cl$_4$N$_{18}$Zn$_5$ | 1.19 | 3.5 | 3 |
| MFU-4*l* | C$_{36}$H$_{12}$Cl$_4$N$_{18}$O$_6$Zn$_5$ | 1.20 / 1.86 | 16.7 | 16 |

MFU-4*l*, a large-pore analogue of MFU-4, contains two different kinds of pores with 1.20 and 1.86 nm diameter, respectively, which are interconnected by 0.9 nm apertures to each other.[42] Due to its large pore apertures, MFU-4*l* can be easily saturated with glycerol, similarly to MFU-1. Despite its low density, interdigitation in MFU-4*l* does not seem to occur, according to experimental evidence gained from previous work.[42]

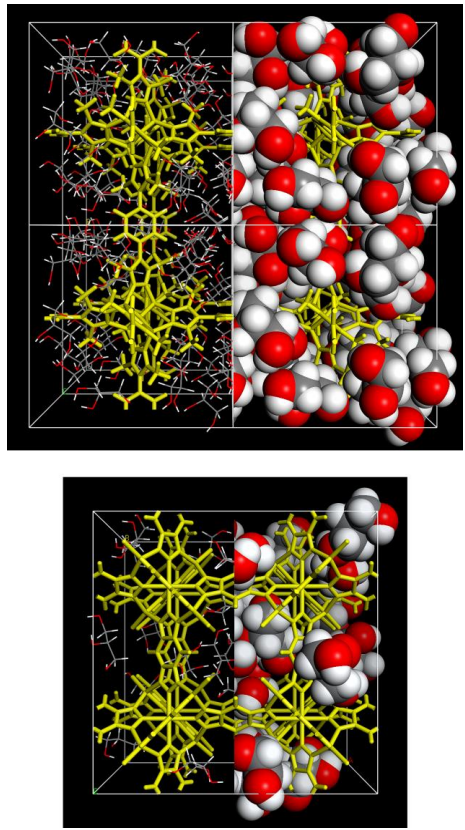

FIG. 1. Representative packing plots of wire models (yellow) of MFU-1 (top) and MFU-4 (bottom), in which internal voids are filled with glycerol. For MFU-1, a 2×2×2 supercell is shown, ensuring that for both MOFs eight secondary building units are visible. Both structures are presented for the same viewing distance.

XRPD measurements show that the glycerol-loaded MOFs retain their crystal structures. The glycerol content in the obtained materials was determined by thermogravimetric analysis (TGA). TGA curves (Fig. S1 in the supplemental material[44]) show that glycerol loss occurs at 100-220 °C (-26.5 %) for MFU-1, 140-260 °C for MFU-4 (-25.7 %) and 100-190 °C (-55.0 %) for MFU-4$l$. As shown in Table I, the experimental average loading of glycerol molecules per formula unit of MFU-4 and MFU-4$l$ matches closely the theoretical values (simulation details are described in section III). It should be noted that for MFU-4, one pore corresponds to two formula units, i.e., one pore contains 7 glycerol molecules. In the case of MFU-4$l$, two formula units correspond to one large (1.86 nm) and one small (1.20 nm) pore. According to simulation, each large pore contains ca. 27 glycerol molecules and each small pore 5 glycerol molecules.

## III. EXPERIMENTAL DETAILS

### A. Preparation and characterization of glycerol in MOF

TGA was performed with a TA Instruments Q500 analyzer in the temperature range of 25–800°C in flowing nitrogen gas at a heating rate of 5 K min$^{-1}$. Powder X-ray diffraction data were collected in the 2θ range of 4–70° with 0.02° steps, with a time of 200 s per step, using a Seifert XRD 3003 TT diffractometer equipped with a Meteor 1D detector.

MFU-1[37] and MFU-4$l$[42] were prepared according to previously described procedures. The samples of MFU-1 and MFU-4$l$ (25 mg) were degassed for 20 h at 200 °C in vacuum and then placed in an open vial into a Schlenk flask containing glycerol (2 ml). The Schlenk flask was heated for 20 h at 60 °C in the vacuum of a rotary pump (approx. 1 mbar). MFU-4 was prepared according to a previously described procedure.[41] The sample of MFU-4 (25 mg) was degassed for 20 h at 320 °C in vacuum and then heated for 20 h at 140 °C with glycerol (5 ml) in a sealed tube. After cooling to room temperature, the mixture was diluted with methanol (10 ml), the precipitate was filtered off, washed with methanol (2×10 ml), and dried in vacuum.

Glycerol-loaded unit cells of all framework compounds were created by the "Sorption Tools" module of Accelrys Materials Studio V7.0, employing a Metropolis sampling scheme to find appropriate positions of the glycerol molecules (loading at 298 K to a fixed target pressure of 100 kPa) in the void volume of MFU-1, MFU-4, and MFU-4$l$, respectively. Saturation was reached in each case after sampling 1×10$^7$ different configurations. During sampling, all framework lattice atoms were fixed at their crystallographic positions. Fig. 1 shows representative low-energy configurations of glycerol-loaded frameworks obtained for the maximum possible loading (for MFU-4$l$, see Fig. S2 in the supplemental material[44]). The numbers of glycerol molecules ($N_{sim}$) per framework formula unit obtained from sorption simulations are given in Table I.

### B. Dielectric measurements

Dielectric spectra of the complex permittivity covering a frequency range of about 10$^{-1}$ Hz - 3 GHz were measured by combining two experimental techniques.[45] A frequency-response analyzer (Novocontrol α-analyzer) was employed in the low-frequency range ($\nu$ < 3 MHz). For the radio-frequency and microwave range (1 MHz < $\nu$ < 3 GHz) a reflectometric technique was used. For these experiments the sample capacitor is mounted at the end of a coaxial line[46] and the measurements are performed using an Agilent E4991A impedance analyzer. For cooling and heating of the samples, a closed-cycle refrigerator, a nitrogen-gas cryostat, and a home-made oven were used.

All dielectric measurements were performed on powder samples to avoid any pressure-induced deterioration of the sample materials that may arise when preparing pellets. The sample powders were filled into parallel-plate capacitors with plate distances between 100 and 150 µm. While slight pressure was applied to the capacitor plates, the obtained absolute values of the measured dielectric permittivity may nevertheless be somewhat reduced due to a limited packing density. The filled capacitors were mounted into the cooling/heating device and kept under vacuum for at least 12 hours before the temperature-dependent dielectric measurements were started. This ensured that residual amounts of water or other contaminations adsorbed on the powder surface were removed.

## IV. RESULTS AND DISCUSSION

As a typical example of the obtained results, Fig. 2 shows spectra of the dielectric constant $\varepsilon'$ and loss $\varepsilon''$ as measured at various temperatures for glycerol confined in MFU-1. For the two other host materials, qualitatively similar spectra were obtained. At high frequencies, the clear signatures of a relaxational process show up: a step in $\varepsilon'(\nu)$ and a peak in $\varepsilon''(\nu)$, both shifting to lower frequencies when the temperature is lowered. This shifting indicates a continuous slowing down of molecular dynamics, typical for glassy freezing.[47,48] We ascribe these spectral features to the structural α-relaxation of confined glycerol. In the spectral region where this relaxational response shows up, $\varepsilon'(\nu)$ and $\varepsilon''(\nu)$ are significantly larger than the results from a measurement of "empty" MFU-1, shown as dashed lines in Fig. 2. This demonstrates that the observed relaxational process indeed arises from the dynamics of glycerol and is not due to the host material.

In addition, for the higher temperatures both $\varepsilon'(\nu)$ and $\varepsilon''(\nu)$ reveal a strong increase at low frequencies. The $\varepsilon''(\nu)$ spectrum at 300 K exhibits the onset of a peak at the lowest frequencies and a slight change of slope between 10 and 100 Hz, indicating further relaxational processes. Generally, measurements of confined supercooled liquids may reveal two additional relaxational processes besides the $\alpha$ relaxation: One may arise from molecules interacting with the pore walls, which usually leads to a slowing down of molecular motion.[11,12,13] The second is expected due to the fact that the host/guest system can be regarded as a highly heterogeneous system.[11,13,21,49] As shown long ago by Maxwell and Wagner, dielectric spectra of systems composed of two dielectric materials can exhibit a non-intrinsic relaxation process.[50,51] It can be completely understood, e.g., within an equivalent-circuit approach, without invoking any frequency-dependent microscopic processes.[52,53]

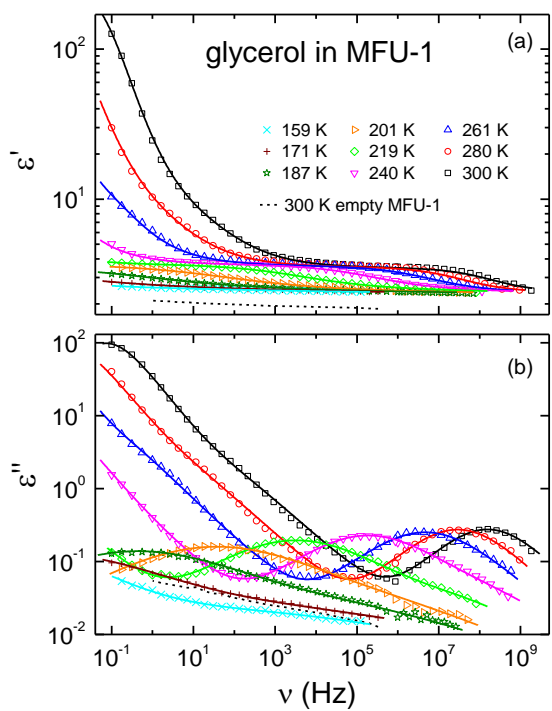

The spectra of Fig. 2 can be well fitted by the sum of three relaxation functions, namely a Cole-Cole (CC) and Havriliak-Negami (HN) function for the mentioned two low-frequency relaxations and a CC function for the $\alpha$ relaxation of confined glycerol (lines in Fig. 2). Both are empirical functions often used to parameterize relaxation phenomena.[47] (For low temperatures, the two low-frequency contributions partly could be omitted in the fits when they were shifted out of the frequency window.) At low temperatures, $T \leq 240$ K, an additional CC function was necessary to account for some excess intensity at the right flanks of the $\alpha$ peaks in the loss, reminiscent of the well-known excess wing observed in glycerol and other glass formers.[48,54,55] However, this spectral feature, which should show up as a second, more shallow power law at the high-frequency flank of the main loss peak, only becomes visible in the spectra for the two lowest plotted temperatures in Fig. 2(b). Here the loss is of similar amplitude as the response of the empty host system (dashed line) and, thus, the significance of this feature is limited and it is not treated in detail here.

Figure 3 compares the dielectric $\alpha$-relaxation peaks of glycerol confined in MFU-1 to bulk glycerol.[48] Obviously, the loss peaks of the confined sample are of much smaller amplitude. When comparing the values of the relaxation strength $\Delta\varepsilon$ obtained from fits, $\Delta\varepsilon$ of the confined sample is by about a factor of 30 smaller than for the bulk. Such a depression of relaxation strength in confined geometry is a common finding.[12,13,16,49] It can be partly ascribed to the trivial fact that the amount of supercooled liquid per volume is reduced in the confined system due to the presence of the host material. However, as pointed out, e.g., in Ref. 49, a simple correction of $\Delta\varepsilon$ for the liquid/host volume ratio is not justified as the different components in a heterogeneous dielectric generally do not combine in an additive way. Moreover, glycerol molecules being slowed down or becoming completely immobile due to interactions with the pore walls also should lead to a reduction of $\Delta\varepsilon$. Finally, the incomplete space filling of the measured powders also leads to a reduction of $\Delta\varepsilon$.

FIG. 2. Frequency dependence of the dielectric constant (a) and loss (b) of glycerol confined in MFU-1, measured at various temperatures. The solid lines are fits, performed simultaneously for $\varepsilon'(\nu)$ and $\varepsilon''(\nu)$, using the sum of a CC and a HN function for the low-frequency response and a CC function for the main relaxation process. For $T \leq 240$ K, an additional CC function was used to account for the excess wing. The dashed lines show the response of the empty host system without glycerol.

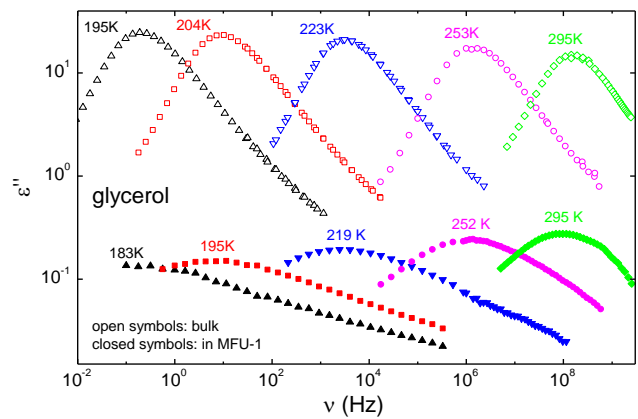

FIG. 3. Frequency dependence of the dielectric loss of bulk glycerol[48] (open symbols) and of glycerol confined in MFU-1 (closed symbols). For the latter, spectra are show at temperatures selected to achieve an approximate match of the peak frequencies with those of the bulk results.

Another effect of confinement revealed by Fig. 3 is a strong broadening of the loss peaks compared to the bulk material, leading to significantly reduced slopes of the low- and high-frequency flanks of the peaks. At room temperature the half width increases from 1.6 to 2.1 decades, at 200 K from 2.1 to 4.4. Such a confinement-induced broadening is a well-known phenomenon[11,12,16,19]. It may be ascribed to interactions of the liquid with the pore walls and/or a variation in the number of glycerol molecules per pore.

In Fig. 3, for the shown spectra in confinement temperatures were selected that lead to comparable peak frequencies as in the bulk data. Comparing the temperatures of both data sets reveals differences that become most pronounced at low temperatures. This signifies the most interesting effect of confinement, namely a significant variation of molecular dynamics, characterized by the relaxation time $\tau$, which is related to the inverse peak frequency via $\tau \approx 1/(2\pi\nu_p)$. It is mainly this shift in $\tau$ and the related variation of the glass-transition temperature that has generated so much interest in confinement measurements of supercooled liquids as it enables conclusions on the role of cooperativity in the glass transition.[12,13,14,15,19,21] In our data, this shift is not caused by a variation of density or negative pressure in confinement[15,16,40] as measurements with different liquid/host ratios did not reveal any shift of $\tau$.

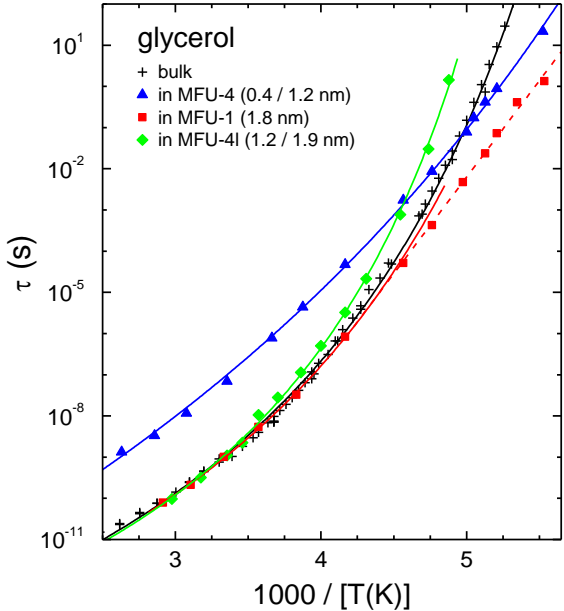

FIG. 4. Arrhenius plot of the $\alpha$-relaxation times of bulk glycerol (plusses) and glycerol confined in three different MOFs. The solid lines are fits with the VFT law (bulk: $\tau_0 = 3.9 \times 10^{-15}$ s, $D = 16$, $T_{VF} = 132$ K; MFU-4: $\tau_0 = 1.8 \times 10^{-15}$ s, $D = 55$, $T_{VF} = 73$ K; MFU-1: $\tau_0 = 4.0 \times 10^{-15}$ s, $D = 16$, $T_{VF} = 130$ K; MFU-4l: $\tau_0 = 5.9 \times 10^{-15}$ s, $D = 12$, $T_{VF} = 149$ K).

The $\tau(T)$ results for the $\alpha$ relaxation of glycerol confined in the three MOFs investigated in the present work are shown in Fig. 4, together with the relaxation times of the bulk material.[48,56] The latter exhibits the well-known deviations from thermally activated Arrhenius behavior, which can be parameterized by the empirical Vogel-Fulcher-Tammann (VFT) law, usually written in the modified form:[57]

$$\tau = \tau_0 \exp\left[\frac{DT_{VF}}{T - T_{VF}}\right] \quad (1)$$

Here $\tau_0$ is an inverse attempt frequency and $T_{VF}$ is the Vogel-Fulcher temperature, where $\tau$ diverges. The so-called strength parameter $D$ is a measure of the deviation from Arrhenius behavior.[57]

For confined glycerol in MFU-1, at temperatures above about 250 K, $\tau(T)$ agrees well with the relaxation time of bulk glycerol (Fig. 4). As noted, e.g., in Ref. 49, Maxwell-Wagner effects arising from the heterogeneous nature of confined samples can lead to a shift of the observed loss peaks to higher frequencies, i.e., an apparent acceleration of the $\alpha$-relaxation. However, the mentioned agreement at high temperatures makes such effects unlikely. Additionally, this shift should become negligible for high liquid/host volume ratios. This indeed is the case for the investigated MOFs where the wall thickness is small compared to the large pore dimensions, which are filled by glycerol. Moreover, we also performed measurements after removal of part of the glycerol molecules by heating in vacuum., i.e., with different liquid/host ratio. This led to no detectable frequency shift of the $\alpha$-relaxation peaks, thus excluding any influence from Maxwell-Wagner effects on $\tau$. The good agreement of $\tau(T)$ for bulk and confined glycerol at high temperatures also indicates that interactions between guest molecules and pore walls play no important role for the $\alpha$ relaxation in this system.

While there is good agreement at high temperatures, $\tau(T)$ of glycerol confined in MFU-1 successively starts to deviate from the bulk curve at low temperatures. As indicated by the dashed line, below about 225 K (1000/$T \approx 4.4$ K$^{-1}$) $\tau(T)$ crosses over from VFT to Arrhenius behavior, showing significantly weaker temperature dependence than the bulk sample. This finding is in full accord with the notion of an increasingly cooperative nature of molecular motions when the glass transition is approached. This leads to a growth of the effective energy barrier for molecular motions explaining the typical deviations from Arrhenius behavior of bulk glass formers.[2,5,6,7,8,10] An increase of molecular cooperativity implies a growing cooperativity length. Within this framework, the mentioned deviation from bulk behavior observed at 225 K can be ascribed to this length exceeding the typical pore size of the host material.[12] Below this temperature, a further increase of cooperativity length is prevented by the confinement. Consequently, the effective energy barrier stops increasing below 225 K and $\tau(T)$ exhibits conventional thermally activated behavior. It seems obvious that, despite the relatively large apertures between the pores in MFU-1, the confinement in MFU-1 is efficient and allows for detecting the growing cooperativity length in glycerol.

The findings discussed above indicate a cooperativity length of glycerol at 225 K of about 1.8 nm, the typical pore size of the host system. However, due to the partial interdigitation of the host framework in MFU-1 (cf. section II), leading to a distribution of pore sizes, and due to the probably only partial filling of the pores, this value should be regarded as a rough estimate only. In any case, a value of $L_{corr} = 1.8$ nm at 225 K in glycerol ($T/T_g \approx 1.20$) would compare reasonably with $L_{corr} = 2.5$ nm at 250 K ($T/T_g \approx 1.15$) deduced from qualitatively similar results on glass forming salol confined in controlled pore glasses.[12]

An alteration of relaxation time compared to the bulk is quite a common finding for confined glass formers.[11,12,13,14,15,16,19] However, in most cases the whole $\tau(T)$ curve becomes shifted in confinement indicating a rather dramatic modification of molecular dynamics, which may at least partly be caused, e.g., by wall interactions or steric hindrance of strongly confined systems. In contrast, a temperature-dependent crossover to Arrhenius behavior triggered by a confinement-induced suppression of a further growth of cooperativity length, until now was only rarely observed.[12,16] It should be noted that the suppression of a further increase of cooperativity at low temperatures also leads to a significant decrease of the glass temperature in the confined system. Using the often-employed condition, $\tau(T_g) \approx 100$ s, we arrive at $T_g \approx 170$ K for glycerol confined in MFU-1 instead of $T_g \approx 188$ K determined in the same way for bulk glycerol. Such a reduction of the glass temperature also is a common finding for confined materials.[12,13,14,15,16,18,21]

The triangles in Fig. 4 show the relaxation-time results for glycerol confined in MFU-4. This host system features two different well-defined pore sizes with average diameters of 0.4 and 1.2 nm.[41] The apertures between the pores have a size of 0.25 nm, not allowing for interpore diffusion of glycerol, and no interdigitation exists in this material. Thus, it represents an ideal host system for confinement measurements. The minor pores are too small to host any glycerol molecules and thus the relevant pore size of this system is 1.2 nm, still significantly smaller than the pores in MFU-1 (1.8 nm). As revealed by Fig. 4, $\tau(T)$ of glycerol confined in this system deviates from the bulk curve in the whole investigated temperature range. Similar behavior has also been reported for several other host/liquid systems.[11,13,14] If assuming that, just as for MFU-1, glycerol-wall interactions play no significant role for this process, these findings imply that, in the whole investigated temperature range, a confinement of 1.2 nm is clearly too small to allow for glassy dynamics as found for the bulk. Obviously, molecular motions and their glassy freezing in this strongly confined system have not much in common with the dynamics found in the bulk supercooled liquid.[15] Notably, the $\tau(T)$ curve for the glycerol/MFU-4 system exhibits much less curvature, i.e., less deviations from thermally activated behavior if compared to the bulk and also to glycerol in MFU-1. This implies lower fragility within the fragile/strong classification scheme of glassy matter[57] (the strength parameter $D$ is 55 instead of 16 in the bulk) and can be interpreted as an indication of less cooperativity. Interestingly, at about 200 K the $\tau(T)$ curve of bulk glycerol crosses the curve of glycerol in MFU-4, i.e., at low temperatures the molecular motions for unconfined glycerol become slower than for the confined system. Apparently, the increasing cooperativity length slows down the bulk dynamics so strongly at low temperatures that the molecular motions become slower than in MFU-4 (and also in MFU-1). In agreement with the findings in other confined supercooled liquids,[12,13,14,15,16,18,21] the glass temperature of this liquid/host system (178 K), determined via $\tau(T_g) \approx 100$ s, is smaller than for the bulk (188 K).

That still some work has to be done to fully understand the behavior of supercooled liquids confined in MOFs is demonstrated by the rather puzzling results on the host system MFU-4*l*. It comprises two types of pores with well-defined sizes of 1.2 and 1.9 nm. Similar to MFU-1, three relaxation processes could be clearly identified in the spectra (see Fig. S3 in the supplemental material). Two of them are many decades slower than the $\alpha$-relaxation of bulk glycerol and most likely arise from Maxwell-Wagner effects and pore-wall interactions. Similar to MFU-1, for glycerol in MFU-4*l* the relaxation time of the remaining process agrees with the bulk behavior at high temperatures (diamonds in Fig. 4). However, already below about 290 K ($1000/T \approx 3.5$ K$^{-1}$), $\tau(T)$ becomes *larger* than that of the bulk, i.e., the dynamics is slowed down due to the confinement. Consequently, here $T_g$ in confinement (198 K) is higher than for bulk glycerol (188 K). In principle, interactions of the glycerol molecules with the wall could explain a slowing down of dynamics but it is not clear why this should play a role below 290 K only. Moreover, the implications of the two different pore sizes in this host material are also not clear. They may naively be expected to lead to two separate relaxation processes at low temperatures, arising from the differently confined glycerol.

## V. SUMMARY AND CONCLUSIONS

In this work, we have checked for the feasibility of using MOFs as host systems for the investigation of glassy dynamics in confined supercooled liquids. Indeed, our dielectric measurements of the molecular glass-former glycerol, confined in three MOFs with different pore sizes, have revealed that these materials are well-suited for this kind of investigation. A variety of confinement effects were found showing up as marked deviations of the dynamic properties from those of the bulk material. The structural $\alpha$-relaxation is well defined in all systems. Compared to bulk glycerol, it exhibits broadening and amplitude reduction as known from other confinement measurements.

Of special interest are the results on the temperature dependence of the relaxation time. For glycerol confined in the well-defined and well-separated pores of MFU-4, $\tau(T)$ strongly departs from the bulk in the complete investigated temperature range. In MFU-4 the glycerol molecules are confined in relatively small pores of 1.2 nm diameter that can host up to 7 molecules. Obviously, for glycerol this number is too small to approach bulk behavior even at the highest covered temperature of 380 K, and thus the correlation length of glycerol always remains larger than 1.2 nm. These findings can be compared to those on ethylene glycol confined in various zeolites, which led to the conclusion that a number of five molecules is insufficient to show dynamics comparable to that of the bulk liquid.[15]

In contrast to MFU-4, for the host materials containing bigger pores (MFU-1 and MFU-4*l*) at least at high temperatures bulk behavior is found. However, at low temperatures deviations are revealed. In case of MFU-1 they show up as a reduction of relaxation time, in good accord with the expected behavior when the correlation length exceeds the pore size. In contrast, in MFU-4*l* a higher and more strongly temperature dependent $\tau(T)$ than in the bulk is found at low temperatures, an unexpected finding which deserves further investigation. Overall, our results demonstrate that MOFs are well-suited host systems for the investigation of glassy dynamics via confinement. It seems a promising approach to perform further investigations in other MOF host systems with different, well-defined pore sizes to clarify the temperature dependence of $L_{corr}$ in glycerol and other glass formers.

**ACKNOWLEDGMENTS**

This work was partly supported by the Deutsche Forschungsgemeinschaft via Research Unit FOR 1394 and by the BMBF via ENREKON.

# MFU-type metal-organic frameworks as host materials of confined supercooled liquids - Supplemental Material

J. K. H. Fischer,[1] P. Sippel,[1] D. Denysenko,[2] P. Lunkenheimer,[1,a)] D. Volkmer,[2] and A. Loidl[1]

[1]*Experimental Physics V, Center for Electronic Correlations and Magnetism, University of Augsburg, 86135 Augsburg, Germany*
[2]*Chair of Solid State and Material Science, Institute of Physics, University of Augsburg, 86135 Augsburg, Germany*

## I. TGA RESULTS

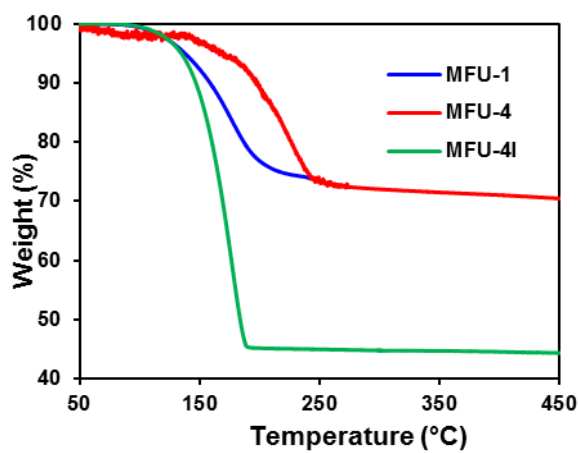

FIG. S1. TGA curves for glycerol@MOF.

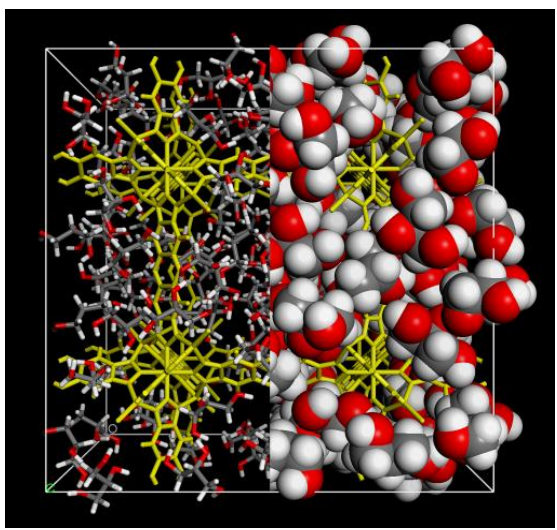

FIG. S2. Representative packing plot of wire model of MFU-4*l* (yellow), in which internal voids are filled with glycerol.

## II. DIELECTRIC SPECTRA FOR GLYCEROL IN MFU-4*l*

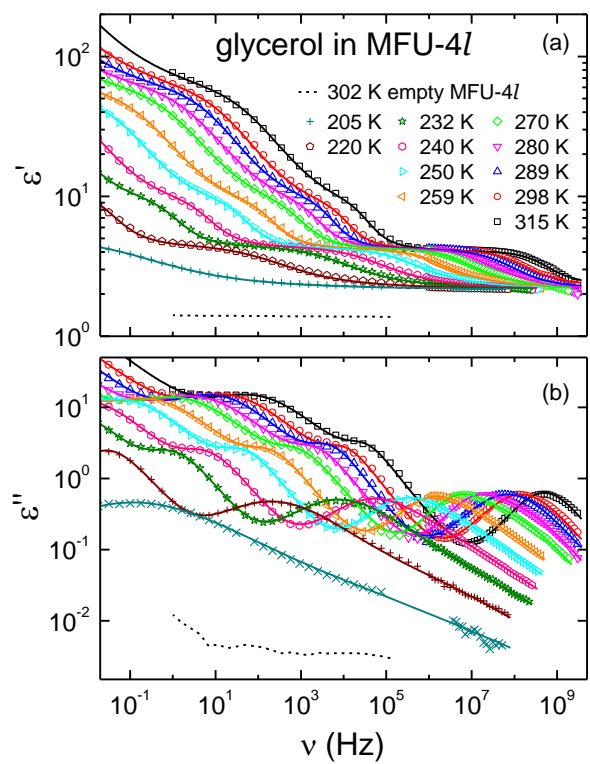

FIG. S3. Frequency dependence of the dielectric constant (a) and loss (b) of glycerol confined in MFU-4*l*, measured at various temperatures. The solid lines are fits, performed simultaneously for $\varepsilon'(\nu)$ and $\varepsilon''(\nu)$, using the sum of up to four CC functions and a power law. The dashed lines show the response of the empty host system without glycerol.

The dielectric spectra of glycerol in MFU-4*l* (Fig. S3) show qualitatively similar behavior as for MFU-1 (Fig. 2 in main paper), i.e., there is a well-defined main relaxation process in a similar frequency/temperature range as for bulk glycerol and additional contributions leading to a strong increase of $\varepsilon'$ and $\varepsilon''$ at low frequencies. In comparison to MFU-1, the two superimposing relaxation processes at lower frequencies, possibly arising from Maxwell-Wagner relaxation and glycerol molecules interacting with the pore walls, become more obvious in MFU-4$l$. To formally account for the additional increase of $\varepsilon'$ and $\varepsilon''$ seen at the lowest frequencies and highest temperatures, an additional power law with exponent $s \approx -0.5$ was used. At the right flank of the well-defined relaxation peaks seen in Fig. S3(b) an additional weak contribution seems to show up, which was taken into account by another CC function. However, its significance is limited.